\documentclass{article}

\usepackage{cite}
\usepackage{graphicx}
\usepackage{epsfig}
\usepackage{amsfonts}
\usepackage{amssymb}
\usepackage{amsmath,amssymb}
\usepackage{color}
\usepackage{epsf,epsfig}
\usepackage{amsmath}
\usepackage{amssymb}
\usepackage{mathrsfs}

\date{\today}

\newcommand{\ee}{\end{equation}}
\newcommand{\eea}{\end{eqnarray}}
\newcommand{\be}{\begin{equation}}
\newcommand{\bea}{\begin{eqnarray}}

\begin{document}

\title{{\bf  
Balancing a static black ring 
\\
with a phantom
scalar field
}}

\author{
{\large Burkhard Kleihaus}$^{1}$,
{\large Jutta Kunz}$^{1}$
and
{\large Eugen Radu}$^{2}$
\vspace{0.5cm}
\\
$^{1}${\small Institut f\"ur  Physik, Universit\"at Oldenburg, Postfach 2503 
  D-26111 Oldenburg, Germany}
   \\
$^{2}${\small Departamento de F\'\i sica da Universidade de Aveiro and CIDMA,}
\\
 {\small Campus de Santiago, 3810-183 Aveiro, Portugal}
} 

  \maketitle  

\small{
\begin{abstract} 
All known five dimensional, asymptotically flat, static 
black rings possess conical singularities.
However, there is no fundamental obstruction 
 forbidding the existence of
balanced configurations, and
we show that
the Einstein--Klein-Gordon equations 
admit (numerical) solutions describing static asymptotically flat 
black rings, which are regular on and outside the event horizon.
The scalar field is 'phantom', 
which
creates the self-repulsion necessary to balance the black rings.
Similar solutions are likely to exist in other spacetime dimensions,
the basic properties of a line element describing a 
four dimensional, asymptotically flat  black ring geometry being
 discussed.

\end{abstract}
}

\section{Introduction and motivation}
 
In 2001
Emparan and Reall
have found  a remarkable new   static, vacuum black
hole (BH) solution  of Einstein equations in $4+1$ dimensions 
\cite{Emparan:2001wk}.
Different from the Schwarzschild-Tangerlini BH \cite{Tangherlini:1963bw}, this
solution has an event horizon with $S^2\times S^1$
topology and describes an  asymptotically flat black ring (BR).
However, the solution in \cite{Emparan:2001wk} is not fully satisfactory,
since it contains a conical singularity 
in the form of a disc ($i.e.$  a negative tension source) that sits inside the ring,
 supporting it against collapse. 
This feature can be understood based on the heuristic construction of a BR starting with a black string
($i.e.$ a four dimensional Schwarzschild BH extending into the fifth dimension)
which is bent to form a circle.
Then, without the tension, this loop would contract, decreasing the radius of $S^1$, due to its 
 gravitational self-attraction\footnote{An analogous construction exists for a special class of {\it non-gravitating}
solitons in four spacetime dimensions -- the vortons,
which are made from loops of vortices, being sustained against collapse by the centrifugal force
\cite{Davis:1988ij}, similar to balanced BRs.}.

Nonvacuum generalizations of the static BR solution  
are known, see $e.g.$
\cite{Kunduri:2004da},
\cite{Emparan:2004wy},
\cite{Kleihaus:2010hd};
however, they still possess conical singularities.
Moreover,
as shown in \cite{Kleihaus:2009dm},
the same result holds also for (static) BRs in Einstein-Gauss-Bonnet theory, 
in which case a region of negative '{\it effective energy density}' (sourced by the Gauss-Bonnet term in the action) occurs.
Although the absolute value of the
conical excess decreases as the Gauss-Bonnet 
coupling constant $\alpha$ increases, the solutions stop to exist
for some $\alpha_{max}$, before approaching a balanced configuration.

So far, the only known mechanism to obtain an asymptotically flat configuration which is
 free of conical singularities
is to set the ring into
rotation \cite{Emparan:2001wn},
in which case the centrifugal force manages to balance the massive ring's self-attraction.

Static, balanced BRs may exist, however, in a non-asymptotically flat background.
For example, as discussed in \cite{Ortaggio:2004kr},
 by submerging a charged
static BR into an electric/magnetic background field, the conical singularities can be
eliminated and the static black ring stabilized. 
However, this construction has the drawback that, 
due to the backreaction of the background electromagnetic field, the BR approaches at infinity
a Melvin-type background.
Although an explicit construction is still missing,
static BRs without conical singularity should also exist in a
 de Sitter spacetime, 
the cosmological expansion acting
against the tension and assuring balance for a critical ring size 
\cite{Caldarelli:2008pz}.
Also, an exact solution describing a static, balanced BR with Kaluza-Klein magnetic monopole asymptotics
has been reported in \cite{Stelea:2012xg}.

 However, there is no fundamental obstruction forbidding the existence of
static, balanced BRs also in a Minkowski spacetime background.
 In fact, such
 line elements can easily be obtained by considering  
(rather mild)
modifications of the Emparan-Reall solution in \cite{Emparan:2001wk}. 
For example,
let us consider the following metric
\begin{eqnarray}
\label{metricBR}
ds^2=\frac{R^2}{(x-y)^2}
\left(
\frac{dx^2}{1-x^2}
+\frac{1+\lambda x}{(y^2-1)(1+\lambda y)}dy^2
+U(x) d\varphi^2
+(1+\lambda x)(y^2-1)d\psi^2
\right)
-\frac{1+\lambda y}{1+\lambda x}dt^2~,
\end{eqnarray}
where $x,y$
are ring coordinates, with the usual range
$
-\infty \leq y \leq -1,
$
$-1 \leq x \leq  1,
$
%
$\varphi$ and $\psi$ are angular directions 
and $t$ is the time coordinate.
Also, $\lambda$ is a free parameter of the solution, with
$
0<\lambda<1,
$
while
$R>0$ is the radius of the ring.
The above line element
possesses an event horizon of $S^2\times S^1$ topology,
located  at 
$
y=- {1}/{\lambda}<-1,
$
 the asymptotic infinity corresponding to $x\to y\to -1$. 
The absence of conical singularities implies that the $\psi$-coordinate
possesses a periodicity
$
\Delta \psi= {2\pi}/{\sqrt{1-\lambda}}.
$
%
The situation is more complicated for the $\varphi$-coordinate,
depending on the choice for the function $U(x)$.
For
\begin{eqnarray}
\label{ch1}
U(x)=(1-x^2)(1+\lambda x)
\end{eqnarray}
one recognizes the static, vacuum Emparan-Reall solution 
\cite{Emparan:2001wk}, 
in which case 
 one cannot eliminate the conical singularities at both $x= -1$ and $x=1$. 
However, no conical singularities are found for particular expressions of the function $U(x)$,
the simplest choice being
\begin{eqnarray}
\label{S}
U(x)= 1-x^2 .
\end{eqnarray}
Then the metric is regular at $x=\pm 1$ (the periodicity of $\varphi$ being $2\pi$),
and, when evaluating various invariant quantities,
  no singularities are found on and outside the horizon, 
while the line element still possesses the proper asymptotic decay.
Moreover, the mass and the Hawking temperature 
 are the same for both (\ref{ch1}) and (\ref{S}), while
 the event horizon area changes accordingly.

However, the vacuum Einstein equations are $not$ solved for the choice (\ref{S}),
the components 
$E_x^x$, 
$E_y^y=E_\psi^\psi$
and
$E_t^t$
of the Einstein tensor being nonzero,
while the expression of the Ricci scalar is
\begin{eqnarray}
{\cal R}=
\frac{3\lambda}{ R^2}\frac{y(1+x^2)-x(1+y^2)}{1+\lambda x}~.
\end{eqnarray}
The Einstein equations are 'satisfied' 
by assuming a matter source with   
$
T_\mu^\nu= E_\mu^\nu/(8\pi G),
$
with $\rho=-T_t^t$
corresponding to the energy density as measured by a fundamental timelike observer.
Then a direct computation  
shows that  $\rho<0$ for some region on and outside the horizon 
(heuristically, this 
provides the repulsive force required for the ring balance). 

Although no field theory source can be associated with the corresponding stress-energy tensor,
the result above suggests that static balanced BRs may exist
  indeed
 in some
models with a matter source violating the weak energy condition. 
The main purpose of this letter is to report on the existence of
such configurations in Einstein gravity
minimally coupled with  a {\it phantom}  real  scalar field.
Such a  field has
 a reverse sign in front of the kinetic energy
part of the Lagrangian density,
which 
leads to the generic occurrence of negative energy densities and  gravitational repulsion.
In the  four dimensional case,
this form of exotic matter has  been considered in cosmology 
and also in wormhole physics, see $e.g.$ 
\cite{Lobo:2005us},
\cite{Kleihaus:2014dla}.
Moreover, 
(spherical)
BH solutions with 'phantom' scalar field hair
do also exist 
\cite{Bronnikov:2005gm},
circumventing
 the no-hair theorems in the
Einstein-scalar field model 
\cite{Herdeiro:2015waa}
due to the violation of the energy conditions. 
Although a phantom scalar 
possesses some undesirable features,
it may perhaps be regarded as corresponding to
 an effective
field theory description
resulting from a fundamental theory which is well defined
\cite{Nojiri:2003vn}
 (see also 	
\cite{Nilles:1983ge}).

For the purposes of this work a phantom scalar  field
is of interest as the simplest source of gravitational repulsion.
Then, our results show that,
for a critical size of the ring, 
this  provides the necessary force
 to keep the BR 
from collapsing,  the resulting configuration being regular,
on and outside the horizon. 
Since no exact solutions are likely to exist in this model,
static balanced BRs are found by solving 
numerically the Einstein--Klein-Gordon equations,
subject to a suitable set of boundary conditions.

This paper is organized as follows. In the next Section
we describe the Einstein-scalar field model. 
For a better understanding of the problem, 
both spherical BHs and BRs are considered.
Then, in Section 3 we  construct the solutions
and show the existence of static, balanced BRs. 
Concluding remarks and some open questions are presented in Section 4.
In particular, an
explicit expression is shown there for 
a four dimensional asymptotically flat BR geometry.

\section{The model }

\subsection{Action, equations and boundary conditions}

We consider the action of a self-interacting real scalar field 
$\phi$ coupled to Einstein gravity in five spacetime dimensions,
\begin{equation}
\label{action}
S=\int  d^5x \sqrt{-g}\left[ \frac{1}{16\pi G} {\cal R}
   -\frac{\epsilon}{2}  g^{\mu\nu} \phi_{, \, \mu}  \phi_{, \, \nu} - V(\phi)
 \right] , 
\end{equation}
where $R$ is the curvature scalar, 
$G$ is Newton's constant,
$V(\phi)$ denotes the scalar field potential,
while 
$\epsilon =1$
for a normal field 
and 
$\epsilon =-1$
for a phantom field.
Using the principle of variation,
one finds the coupled Einstein--Klein-Gordon equations
\begin{equation}
\label{Einstein-eqs}
E_{\mu\nu}=R_{\mu\nu}-\frac{1}{2}g_{\mu\nu} {\cal R}- 8 \pi G~T_{\mu\nu}=0,
~~\frac{1}{\sqrt{-g}} \partial_\mu \big(\sqrt{-g} \partial^\mu\phi \big)= \epsilon \frac{\partial V}{\partial \phi },
\end{equation}  
where $T_{\mu\nu}$ is the
 stress-energy tensor  of the scalar field
\begin{eqnarray}
\label{tmunu} 
T_{\mu \nu}  
=
\epsilon
 \phi_{, \, \mu} \phi_{, \, \nu} 
-g_{\mu\nu} \left[
 \frac{\epsilon}{2}  g^{\alpha\beta} \phi_{,  \alpha}  \phi_{, \beta} + V(\phi)
\right]
 \ .
\end{eqnarray}
 %

{
The solutions in this work are 
 static and axisymmetric,
 with a symmetry group ${\bf R}\times U(1)\times  U(1)$ (where ${\bf R}$ denotes the time translation)
and can be studied 
by using a metric Ansatz
introduced in \cite{Kleihaus:2010pr}, 
with\footnote{Although one can write an Ansatz based on the 
ring coordinates $(x,y)$,
 (which results in a much simpler  
form of the vacuum solution), its use in numerics is problematic, 
at least for the scheme employed in this work,
the asymptotic infinity being approached at a single point.}
\begin{eqnarray}
\label{metric}
 ds^2=f_1(r,\theta)(dr^2+r^2 d\theta^2)+f_2(r,\theta) d\psi^2+f_3(r,\theta) d\varphi^2-f_0(r,\theta) dt^2,
\end{eqnarray}
where the range of $\theta$ is $0\leq\theta\leq \pi/2$ and  
  with $0\leq (\psi,\varphi) \leq 2 \pi$. 
	Also,  $r$ and $t$ correspond  to the radial and time
coordinates, respectively. 
The range of $r$ is $0<r_H\leq r<\infty$ (with $r_H$ the event horizon radius);
thus the $(r,\theta)$ coordinates have a rectangular boundary well suited for numerics.
The scalar field is also a function of $(r,\theta)$, only.

An appropriate combination 
of the Einstein equations,
 $E_t^t=0,~E_r^r+E_\theta^\theta=0$, $E_{\psi}^{\psi}=0$,
 and  $E_{\varphi}^{\varphi}=0$,
yields the following set of equations for the functions $f_1,~f_2,~f_3$ and $f_0$
(where we define
$
(\nabla U) \cdot (\nabla W)=\partial_r U \partial_r W+ \frac{1}{r^2}\partial_\theta U \partial_\theta W,
$
and
$
\nabla^2 U=\partial_r^2U+\frac{1}{r^2}\partial_\theta^2 U+\frac{1}{r}\partial_r U
$):
\begin{eqnarray}
\nonumber
&&
\nabla^2 f_0
-\frac{1}{2f_0}(\nabla f_0)^2
+\frac{1}{2f_2}(\nabla f_0)\cdot( \nabla f_2)
+\frac{1}{2f_3}(\nabla f_0)\cdot( \nabla f_3)
+\frac{32 \pi G}{3}f_0 f_1 V(\phi) =0,
\\
\nonumber
&&\nabla^2 f_1
-\frac{1}{f_1}(\nabla f_1)^2 
-\frac{f_1}{2f_0f_2}(\nabla f_0)\cdot( \nabla f_2)
-\frac{ f_1}{2f_0f_3}(\nabla f_0)\cdot( \nabla f_3)
-\frac{ f_1}{2f_2f_3}(\nabla f_2)\cdot( \nabla f_3) 
\\
\nonumber
&&
{~~~~~~~~~~~~~~~~~~~~~~~~~~~~~~~~~~~~~~}
+8 \pi G f_1 \left(
\epsilon (\nabla \phi)^2-\frac{2f_1}{3}V(\phi)
\right)
=0,
\\
\label{eqs1} 
&&\nabla^2 f_2
-\frac{1}{2f_2}(\nabla f_2)^2
+\frac{1}{2f_0}(\nabla f_0)\cdot( \nabla f_2)
+\frac{1}{2f_3}(\nabla f_2)\cdot( \nabla f_3)
+\frac{32\pi G}{3}f_1 f_2 V(\phi)=0,
\\
\nonumber
&&\nabla^2 f_3
-\frac{1}{2f_3}(\nabla f_3)^2
+\frac{1}{2f_0}(\nabla f_0)\cdot( \nabla f_3)
+\frac{1}{2f_2}(\nabla f_2)\cdot( \nabla f_3) 
+\frac{32\pi G}{3}f_1 f_3 V(\phi) =0~,
\end{eqnarray}
while the Klein-Gordon equation is
\begin{eqnarray}
\label{eqs2} 
\nabla^2 \phi
+\frac{1}{2f_0} (\nabla f_0)\cdot( \nabla \phi)
+\frac{1}{2f_2} (\nabla f_2)\cdot( \nabla \phi)
+\frac{1}{2f_3} (\nabla f_3)\cdot( \nabla \phi)
-\epsilon  \frac{\partial V(\phi)}{\partial \phi}=0.
\end{eqnarray}
%

The remaining Einstein equations $E_\theta^r=0,~E_r^r-E_\theta^\theta=0$
yield two constraints. However, following \cite{Wiseman:2002zc},  
one can show that they are satisfied as well, subject to the boundary conditions given below.

Both BHs with a spherical horizon topology and BRs can be described within the Ansatz (\ref{metric}).
In the vacuum case $(\phi=V(\phi)=0)$, 
the simplest solution is the (spherical) Schwarzschild-Tangerlini  BH \cite{Tangherlini:1963bw}
written in isotropic coordinates, with
\begin{eqnarray}
\label{ST}
f_0=\frac{\big(1-\frac{r_H^2}{r^2}\big)^2}{\big(1+\frac{r_H^2}{r^2}\big)^2},~~
f_1=\frac{f_2}{r^2\cos^2 \theta}=\frac{f_3}{r^2\sin^2 \theta}=\big (1+\frac{r_H^2}{r^2} \big)^2~.
\end{eqnarray}
The corresponding expressions for the 
 (static) Emparan-Reall solution are more complicated, with 
\begin{eqnarray}
\label{BR-vacuum}
&&
f_0
=  \frac{\big(1-\frac{r_H^2}{r^2}\big)^2}{\big(1+\frac{r_H^2}{r^2}\big)^2},~~
f_1
= \frac{\big(1+\frac{r_H^2}{r^2}\big)^2}{\big(1+\frac{r_H^2}{R^2} \big)^2 P }
\left(
\big(1+\frac{r_H^4}{r^4}\big) \big(1+\frac{r_H^4}{R^4}\big) -\frac{4r_H^4}{r^2 R^2}\cos 2\theta +\frac{2r_H^2}{R^2}P 
\right) ,
\nonumber
\\
 &&
f_2=\frac{1}{4f_3}r^4\big(1+\frac{r_H^2}{r^2}\big)^4 \sin^22\theta,
~~
 f_3=
\frac{r^2}{2}
\bigg(
P+ \frac{R^2}{r^2}
(
1+\frac{r_H^4}{R^4}-\frac{r_H^2}{R^2}(\frac{r^2}{r_H^2}+\frac{r_H^2}{r^2})\cos 2 \theta
)  
\bigg)
 \ ,
\nonumber
\end{eqnarray}
where
\begin{equation}
P= \frac{r^2}{2}\left[
\left(1+\left(\frac{R}{r}\right)^4
         -2 \cos 2\theta \left(\frac{R}{r}\right)^2\right)
\left(1+\left(\frac{r_H^2}{r R}\right)^4
         -2\cos 2\theta  \left(\frac{r_H^2}{r R}\right)^2\right)	 
\right]^{1/2} \ ,
\nonumber
\end{equation}
with $R>r_H$ a new parameter, the radius of the ring.
Also,
one can verify that
 the spherical solution (\ref{ST}) is approached as $R\to r_H$. 
Further properties of the static BR for the above parametrization, including
the correspondence with the
Weyl coordinates,
 can be found in  Refs. 
\cite{Kleihaus:2009dm},
\cite{Kleihaus:2010pr}.

 {\small \hspace*{3.cm}{\it  } }
\begin{figure}[t!]
\hbox to\linewidth{\hss%
	\resizebox{9cm}{7cm}{\includegraphics{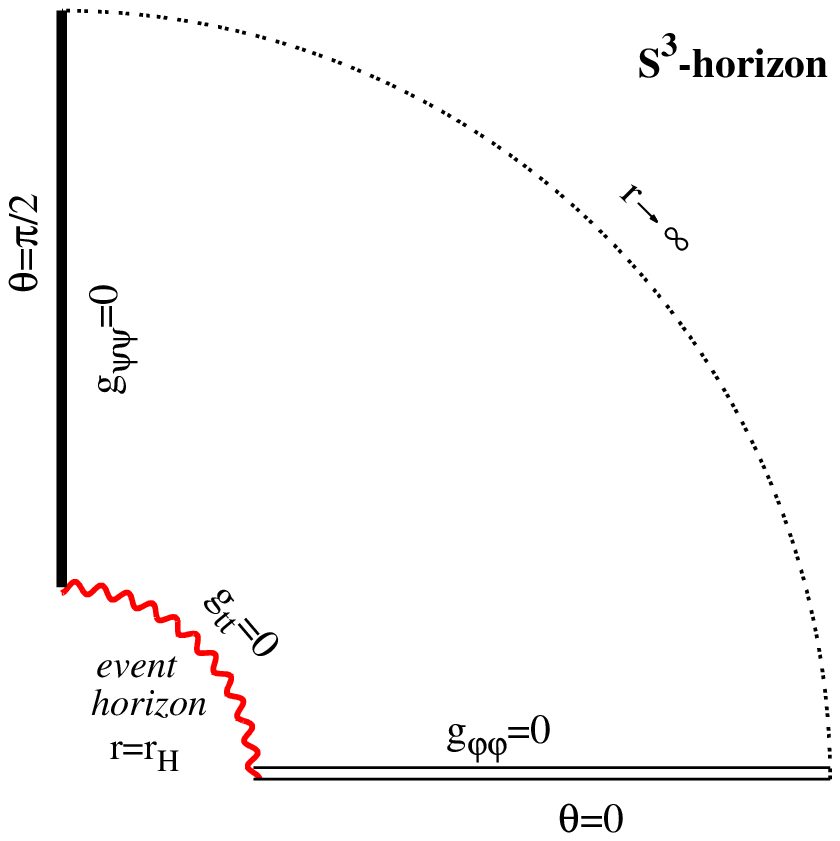}}
	\resizebox{9cm}{7cm}{\includegraphics{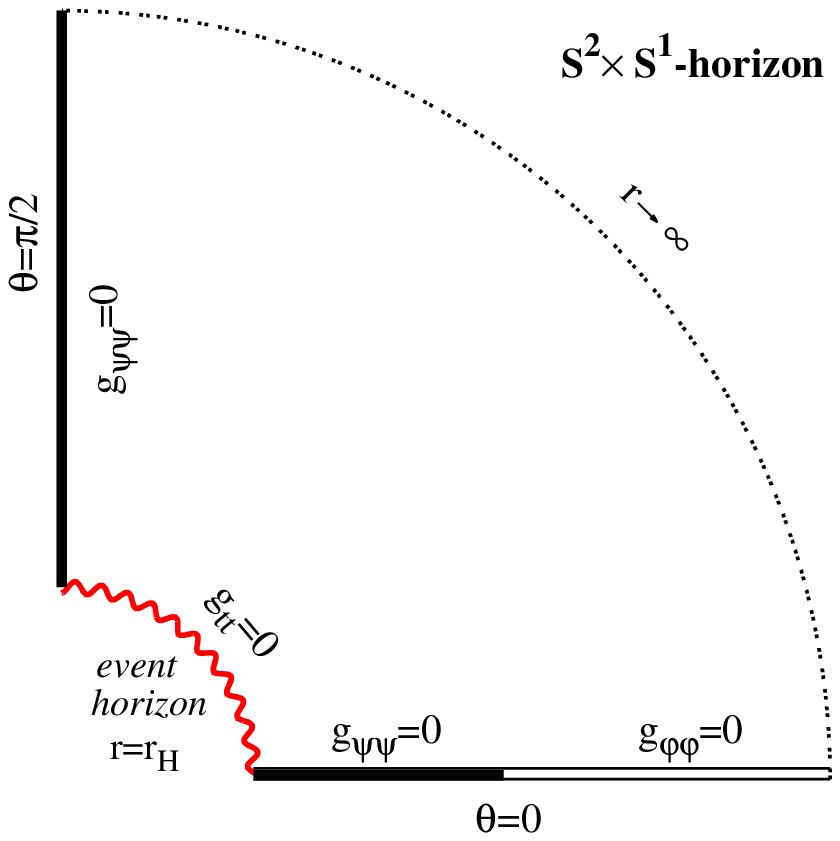}}
\hss}
\caption{\small 
The domain of integration for the
 coordinate system (\ref{metric}) 
is shown for a
 spherically symmetric  black hole (left) 
and a static black ring (right).
}
\label{domain}
\end{figure}

The solutions with $\phi\neq 0$ are found numerically, by
solving the
equations (\ref{eqs1}) 
subject to a set of boundary conditions
which results from the requirement that the solutions describe asymptotically flat
black objects with a regular horizon\footnote{The imposed boundary conditions
(\ref{bc-inf})-(\ref{bc-01})
  are also compatible with 
	an approximate form of the solutions on the boundaries of the domain of integration.
	This domain is shown in Fig. \ref{domain},
	together with the boundary conditions which determine the horizon topology.}.
We assume that as $r\to \infty$,  the Minkowski spacetime background 
(with $ds^2=dr^2+r^2(d\theta^2+\cos^2\theta d\psi^2+\sin^2 \theta d\varphi^2)-dt^2$)
is recovered, while the scalar field vanishes. This implies 
\begin{eqnarray}
\label{bc-inf}
f_0|_{r=\infty}=1,~ f_1|_{r=\infty}=1,~ \lim_{r\to \infty} \frac{f_2}{r^2}=\cos^2\theta,
\lim_{r\to \infty} \frac{f_3}{r^2}=\sin^2\theta,~ \phi|_{r=\infty}=0.
\end{eqnarray}
Also, we impose the existence of a nonextremal event horizon, which is located
 at a constant value of the radial
coordinate, $r=r_H>0$.
There we require 
\begin{eqnarray}
\label{bc-rh}
f_0|_{r=r_H}=0,~\partial_r f_1|_{r=r_H}=\partial_r f_2|_{r=r_H}=\partial_r f_3|_{r=r_H}=0,~~
\partial_r \phi |_{r=r_H}=0.
\end{eqnarray}
The boundary conditions at $\theta=\pi/2$ are
\begin{eqnarray}
\label{bc-pi2}
\partial_\theta f_0|_{\theta=\pi/2}
=\partial_\theta f_1|_{\theta=\pi/2}
= f_2|_{\theta=\pi/2}
=\partial_\theta f_3|_{\theta=\pi/2}= 0,~\partial_\theta  \phi|_{\theta=\pi/2}=0.
\end{eqnarray}
The absence of conical singularities requires  also
$r^2 f_1=f_2$ on that boundary.

The boundary conditions at $\theta=0$ 
are more complicated.
First, for a spherical BH one imposes
\begin{eqnarray}
\label{bc-02}
\partial_\theta f_0|_{\theta=0}
=\partial_\theta f_1|_{\theta=0}
=\partial_\theta f_2|_{\theta=0}
=f_3|_{\theta=0}=0,~~
\partial_\theta \phi |_{\theta=0}
=0.
\end{eqnarray}
For a BR,  a new input parameter, $R>r_H$,
occurs, as for the vacuum solution.
%
There, for $r_H<r<R$, we impose 
\begin{eqnarray}
\label{bc-01}
\partial_\theta f_0|_{\theta=0}
=\partial_\theta f_1|_{\theta=0}
=f_2|_{\theta=0}
=\partial_\theta  f_3|_{\theta=0}= 0,
~~ \partial_\theta  \phi |_{\theta=0}=0.
\end{eqnarray} 

\subsection{Physical quantities}
For any event horizon topology,
the  metric of a spatial cross-section of the horizon is
\begin{eqnarray}
\label{horizon-metric}
d\sigma^2=f_1(r_H,\theta)r^2_H d\theta^2+f_2(r_H,\theta)d\psi^2+f_3(r_H,\theta)d\varphi^2.
\end{eqnarray}
As we shall see, a spherical BH has 
$f_1(r_H,\theta)=f_{10}$, 
$f_2(r_H,\theta)=f_{10}r_H^2 \cos^2\theta$,
$f_3(r_H,\theta)=f_{10}r_H^2 \sin^2\theta$,
such that (\ref{horizon-metric}) parametrizes a round $S^3$.
For a BR,  the orbits of $\psi$ shrink  to zero at $\theta=0$ and $\theta=\pi/2$,
 while 
the length of $S^1$-circle does not vanish anywhere, such that 
the topology of the horizon 
is $S^2\times S^1$ 
(in fact,   $f_2(r_H,\theta)\sim \sin^2 2\theta$
while $f_1(r_H,\theta)$ and $f_3(r_H,\theta)$ are strictly positive and finite functions).
Also, we mention that
although the constants $(R,r_H)$ have no invariant meaning,
they provide a rough measure for the radii of the $S^1$  and $S^{2}$ parts in the horizon metric (\ref{horizon-metric}). 

For both  BRs and spherical BHs, 
the
event horizon area and the Hawking temperature\footnote{The constraint equation $E_r^\theta=0$ guarantees that 
the Hawking temperature $T_H$ is a constant.} are given by
\begin{eqnarray}
\label{AH}
A_H=4\pi^2 r_H \int_{0}^{\pi/2} d\theta \sqrt{f_1 f_2f_3 }\bigg|_{r=r_H}~,~~
T_H= 
\frac{1}{2 \pi }\lim_{r\to r_H}\sqrt{\frac{f_0}{(r-r_H)^2f_1}} ~.
\end{eqnarray} 

At infinity, the Minkowski background is approached.
The ADM mass $M$ of the solutions 
can be read from the asymptotic
expression for the metric function $f_0$,
\begin{eqnarray}
\label{gtt}
-g_{tt}=f_0\sim 1-\frac{8  G M}{3\pi  }\frac{1}{r^{2}}+\dots~.
\end{eqnarray} 
As usual, $M$ can be expressed as the sum of the horizon mass and 
the mass  
stored in the matter field(s) outside the horizon,
which results in the  
  Smarr-type relation
\begin{eqnarray}
\label{smarr}
 M=\frac{3}{2} T_H \frac{1}{4 G}A_H+ M_{(\phi)},
\end{eqnarray}
with
\begin{eqnarray}
\label{smarr1}
 M_{(\phi)}= \frac{3 }{2}\int_{\Sigma} d^4 x\sqrt{-g}\left( \frac{1}{3}T_\nu^\nu- T_t^t \right)=
 -4\pi^2  \int_{r_H}^\infty dr \int_0^{\pi/2} d\theta~
r f_1\sqrt{f_0 f_2 f_3}V(\phi)
,
\end{eqnarray}
(where one integrates over a  spacelike surface $\Sigma$ bounded by 
 the (spatial section of the) horizon and infinity).
Also, we define the  
reduced dimensionless quantities, obtained by dividing out an appropriate power of $M$
\begin{eqnarray}
\label{red}
 a_H=\frac{3}{32}\sqrt{\frac{3}{2\pi}}\frac{A_H}{(G M)^{3/2}},~~t_H=4\sqrt{\frac{2\pi}{3}} T_H\sqrt{GM},
\end{eqnarray}
such that $a_H=t_H=1$
for the Schwarzschild-Tangerlini solution
and 
$a_H=1/t_H=2R r_H/(r_H^2+R^2)$ 
for the Emparan-Reall static BR.

\subsection{The potential, scaling properties and numerics}
 
%
For a quantitative study of the solutions, we need to specify the expression
of the potential $V(\phi)$.
For any horizon topology,
$V(\phi)$ should satisfy the following  relation 
\begin{eqnarray}
\label{rel1}
 \int_\Sigma d^4 x \sqrt{-g} 
\left(
\phi \nabla^2 \phi-\epsilon  \phi \frac{\partial V(\phi)}{\partial \phi}
\right)=0~,
\end{eqnarray}
which is found by
multiplying the Klein-Gordon equation by $\phi $ and integrating it,
 the contribution of the boundary terms vanishing for static, regular solutions
 (with a scalar field that falls off sufficiently fast at infinity).  
This implies that $\phi {\partial V }/{\partial \phi}$
necessarily changes the sign outside the horizon
and rules out a massless (or non-selfinteracting)  field.
 
The results reported in this work 
correspond 
to the simplest polynomial potential which is compatible with
 (\ref{rel1}); we also impose the discrete symmetry of the model
$\phi \to -\phi$.
Thus, for both normal and phantom fields,
 $V$
is taken as the sum of a quadratic and a quartic term, 
\begin{eqnarray}
\label{pot}
V(\phi) =\epsilon ( \frac{1}{2}\mu^2  \phi^2 -\frac{1}{4}\lambda  \phi^4)~.
\end{eqnarray} 
The first term (with  $\mu^2>0$)
provides a  
mass for the scalar field (and leads to an exponential decay 
of the scalar field), 
while  $\lambda$ is a  positive parameter, as required by (\ref{rel1}). 
 
With the above choice of the potential, the system possesses two scaling symmetries 
(with $c$ some positive constant)
 \begin{eqnarray}
 \label{s1} 
(i)~~  r \to  r c,~~\mu\to \mu/c,~\lambda\to \lambda/c^2,~~ 
 {\rm and}~~~
 (ii)~~\phi\to \phi c,~~\lambda \to \lambda/c^2,~~G\to G/c^2 ~,
 \end{eqnarray} 
 which are used to
set to one the values of the 
constants $\mu$ and $\lambda$.
%
This reveals the existence of the dimensionless parameter
 \begin{eqnarray}
\alpha^2=\frac{4\pi G \mu^2}{\lambda}
 \end{eqnarray} 
 characterizing a given model.\footnote{
 Thus the Einstein equations 
solved numerically are
$R_{\mu\nu}-\frac{1}{2}g_{\mu\nu} {\cal R}= 2\alpha^2~T_{\mu\nu}$.
}
%

The BRs are found by employing a finite difference solver \cite{schoen},
which uses  a Newton-Raphson method.
We also mention that
 the required boundary behaviour of the metric functions is
enforced by taking
 %
$
 {f}_i= f_i^{(0)}{F}_i ,
$
where the background functions $ {f}_i^{(0)}$ 
are those of the vacuum BR as given by (\ref{BR-vacuum}).
The advantage of this approach is that the coordinate singularities
are essentially subtracted, while imposing at the same time the $S^2\times S^1$ 
event horizon topology.
Then the numerics is done in terms of the new functions 
$F_i$, subject to
  a set of 
  boundary conditions which follows directly from 
(\ref{bc-inf})-(\ref{bc-01})
together with (\ref{BR-vacuum}).
In the spherically symmetric case, 
 the equations  are solved by using a standard Runge-Kutta  solver and implementing a shooting method.

Let us mention that 
the formalism described above 
holds
for both values of
$\epsilon$.
Also,
we have considered solutions of the equations 
(\ref{eqs1}),  (\ref{eqs2})  
with $\epsilon =\pm 1$. 
However, he have failed to find
balanced BR solutions with a normal   
scalar field 
(despite the 
occurrence of negative energy densities also in that case).
Therefore, for the remainder   of this work
we shall consider the case of a phantom field only, 
$\epsilon=-1$.

 {\small \hspace*{3.cm}{\it  } }
\begin{figure}[t!]
\hbox to\linewidth{\hss%
	\resizebox{9cm}{7cm}{\includegraphics{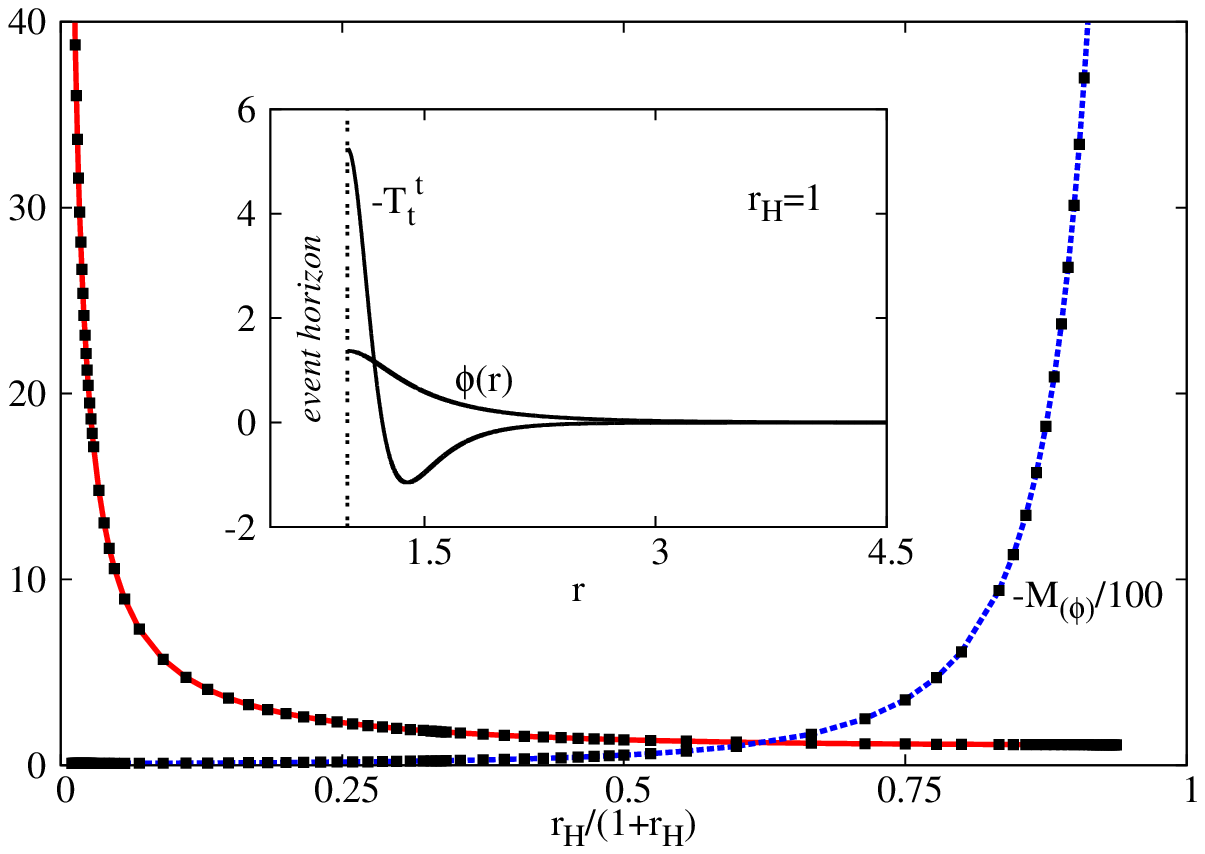}}
	\resizebox{9cm}{7cm}{\includegraphics{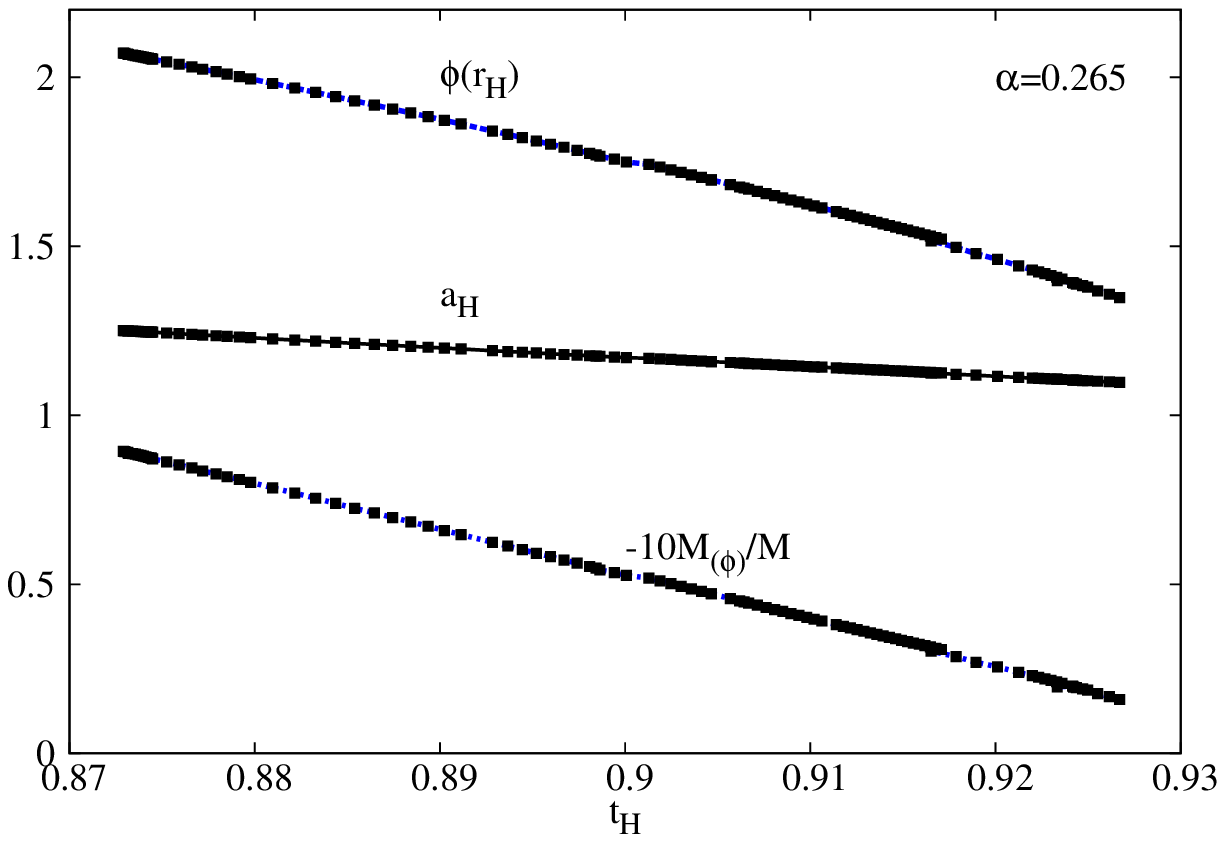}}
\hss}
\caption{\small 
{\it Left.} The value of the scalar field
at the horizon $\phi(r_H)$ and the  mass  $M_{(\phi)}$
are shown for solutions
in  a fixed Schwarzschild-Tangerlini background with a given event horizon radius $r_H$.
The inset shows the profile of a typical (non-gravitating)
solution.
{\it Right.} Some parameters of spherically symmetric black holes with phantom
scalar hair are shown as a function of (reduced) temperature 
for a fixed value of the coupling constant $\alpha$.
Note, that in all figures in this work
exhibiting results for families of solutions, the large dots represent the data points.
}
\label{BH-spherical}
\end{figure}

\section{The solutions}

\subsection{Spherically symmetric black holes}
Let us start with a discussion of the spherically symmetric gravitating solutions.
These configurations are easier to construct,
while their study 
 helps in understanding
some of the BRs properties.

 In this case,
the scalar field is a function of $r$ only,  
while the metric ansatz simplifies, with a factorized angular dependence
  \begin{eqnarray}
 \label{sph1}
f_2 = f_1 r^2 \cos^2 \theta,~~f_3 = f_1 r^2 \sin^2 \theta,~~
  \end{eqnarray} 
	while $f_0$, $f_1$ depend on $r$ only.
The horizon of the black holes is located at $r=r_H>0$, where the solutions have a power-series expansion (for completeness,
here we restore the proper factors of $\mu$, $\lambda$):
\begin{eqnarray} 
\label{sol-hor}
\phi(r)=\phi_0+\frac{1}{4}f_{10}\phi_0(\mu^2-\lambda \phi_0^2)(r-r_H)^2+\dots,
f_0(r)= f_{02}(r-r_H)^2-\frac{f_{02}}{r_H}(r-r_H)^2+\dots, ~
\\
\nonumber
f_1(r)=f_{10} -\frac{2f_{12}}{r_H}(r-r_H) 
 +f_{10}(
4-\frac{ 1}{2} \alpha^2f_{10}r_H^2 \phi_0^2(\mu^2-\frac{1}{2}\lambda \phi_0^2)
)(r-r_H)^2 +\dots, ~
\end{eqnarray}
 in terms of three parameters  $f_{10}=f_1(r_H)$, $f_{02}=f_0''(r_H)/2$ and  $\phi_0=\phi(r_H)$.
One can write an approximate form of the solutions also for $r\to \infty$, with
\begin{eqnarray} 
\label{sol-inf}
f_0(r)=1+\frac{\bar f_{02}}{r^2}+\frac{\bar f_{02}^2}{2r^4}+\dots, ~~
f_1(r)=1-\frac{\bar f_{02}}{2r^2}+\frac{\bar f_{02}^2}{16r^4}+\dots, ~~
\phi(r)=\bar \phi_1 \frac{e^{-\mu r}}{r^(3/2)}+\dots,
\end{eqnarray}
with $\bar f_{02},\bar \phi$ two constants fixed by numerics\footnote{Note that
only nodeless scalar field configurations are reported here (including the BR case).
However, excited solutions  do also exist.
}.

In the study of these solutions, it is useful to consider first
the solutions 
of the Klein-Gordon equation (\ref{eqs2})
in
a fixed BH background as given by the Schwarzschild-Tangerlini metric (\ref{ST}),
$i.e.$
the probe limit, $\alpha=0$.
The corresponding equation reads
\begin{eqnarray} 
\label{probe1}
 \phi''
+\bigg(\frac{3+\frac{r_H^4}{r^4}}{1-\frac{r_H^4}{r^4}} \bigg)\frac{\phi'}{r}
-(1+\frac{r_H^2}{r^2})^2(\mu^2-\lambda \phi^2)\phi=0.
\end{eqnarray}
As seen in Figure \ref{BH-spherical} (left panel),
the solutions  exist for very large (possible arbitrarily large) values of $r_H>0$.
However, the Minkowski spacetime limit $r_H \to0$ is not well defined, 
with a divergent scalar field\footnote{This results from  the virial 
identity 
$T+2\mu^2 V_1-\lambda V_2=0$,
with the strictly positive quantities
$T=\int_0^\infty dr r^3 \phi'^2$,
$V_1= \int_0^\infty dr r^3 \phi^2$,
$V_2=  \int_0^\infty dr r^3 \phi^4$.
Since the Bekenstein-type relation 
(\ref{rel1})
implies
$T+\mu^2 V_1-\lambda V_2=0$,
one finds $V_1=0$, and thus $\phi=0$.
 }.
Also, the mass of these configurations, $M_{(\phi)}=-\int_\Sigma d^4x \sqrt{-g}  T_t^t $, is always negative.

Including the backreaction leads to a fundamental branch of solutions
describing BHs with scalar hair.
As expected,
the solutions with a given horizon size exist for a finite range of $\alpha$.
Moreover, for given $\alpha$,
more than one solution with the same value of $r_H$
(or even the same horizon size) 
may exist.
This can be understood by noticing that the limit $\alpha \to 0$
can be approached as $G\to 0$ ($i.e.$ no backreaction, a fixed BH background)
or as $\mu\to 0$ 
(which corresponds to a model with a massless scalar field).
Moreover, these branches are not always connected.
Also, we mention that BHs  with  $M<0$ 
are also found,
in which case the mass stored in the scalar field  
$M_{(\phi)}$ dominates in (\ref{smarr}) over the horizon mass
(typically found on the  branch
connected with the $G \to 0$ limit). 

In  Figure \ref{BH-spherical} (right panel)
we show some properties of the  solutions with a given $\alpha$,
as a function of the scaled temperature $t_H$.
As $t_H\to t_H^{(min)}$, the numerics becomes increasingly difficult,
a singular solution being approached, with
a divergent Kretschmann scalar as $r\to r_H$.
No singularities are found as $t_H\to t_H^{(max)}$,
in which limit the solutions seem  to continue into
a branch of wormhole configurations.
A systematic discussion of the spherically symmetric 
solutions with $\epsilon=\pm 1$ will be presented elsewhere.

 {\small \hspace*{3.cm}{\it  } }
\begin{figure}[t!]
\hbox to\linewidth{\hss%
	\resizebox{9cm}{7cm}{\includegraphics{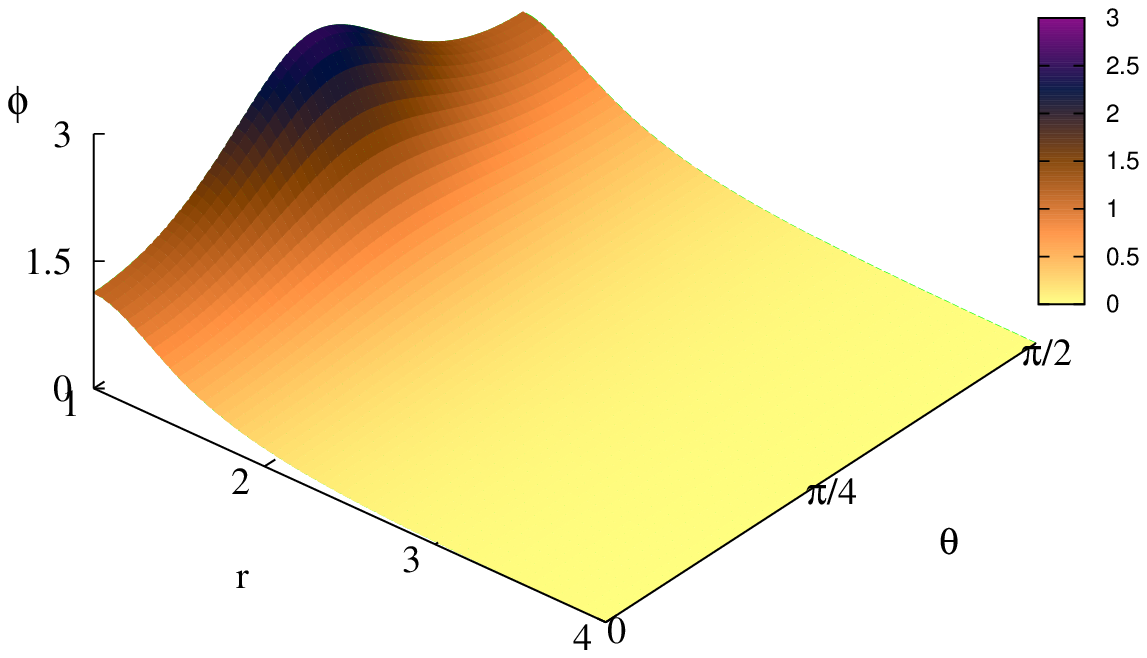}}
	\resizebox{9cm}{7cm}{\includegraphics{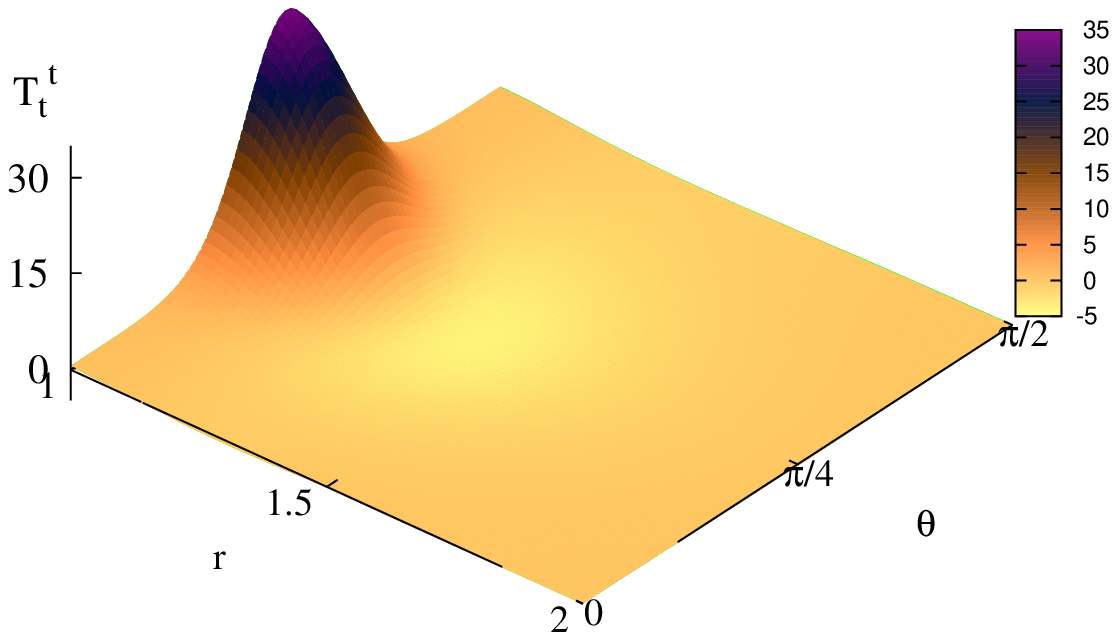}}
\hss}
\caption{\small 
The profile of a typical (non-gravitating) solution in a fixed black ring background with
$r_H=1$, $R=2.$
}
\label{probe-BR}
\end{figure}

 {\small \hspace*{3.cm}{\it  } }
\begin{figure}[t!]
\hbox to\linewidth{\hss%
	\resizebox{9cm}{7cm}{\includegraphics{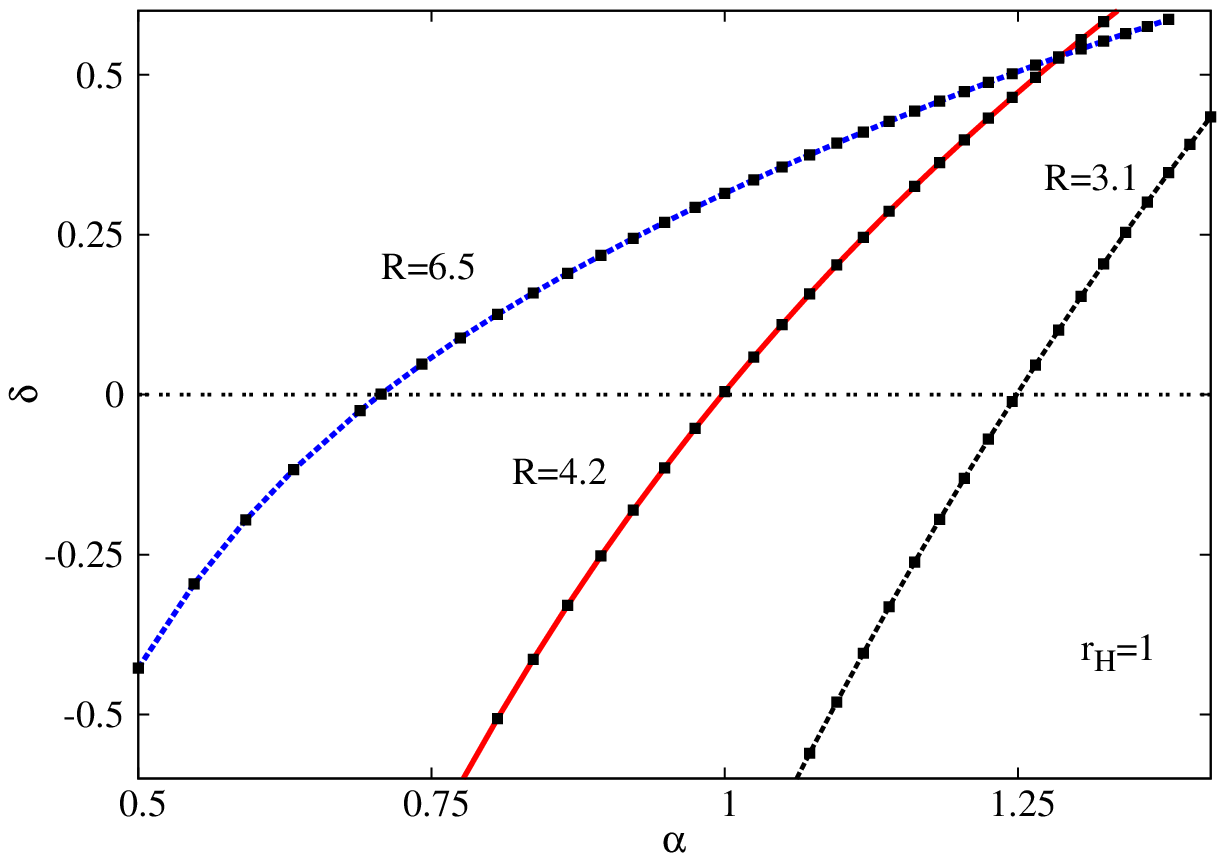}}
	\resizebox{9cm}{7cm}{\includegraphics{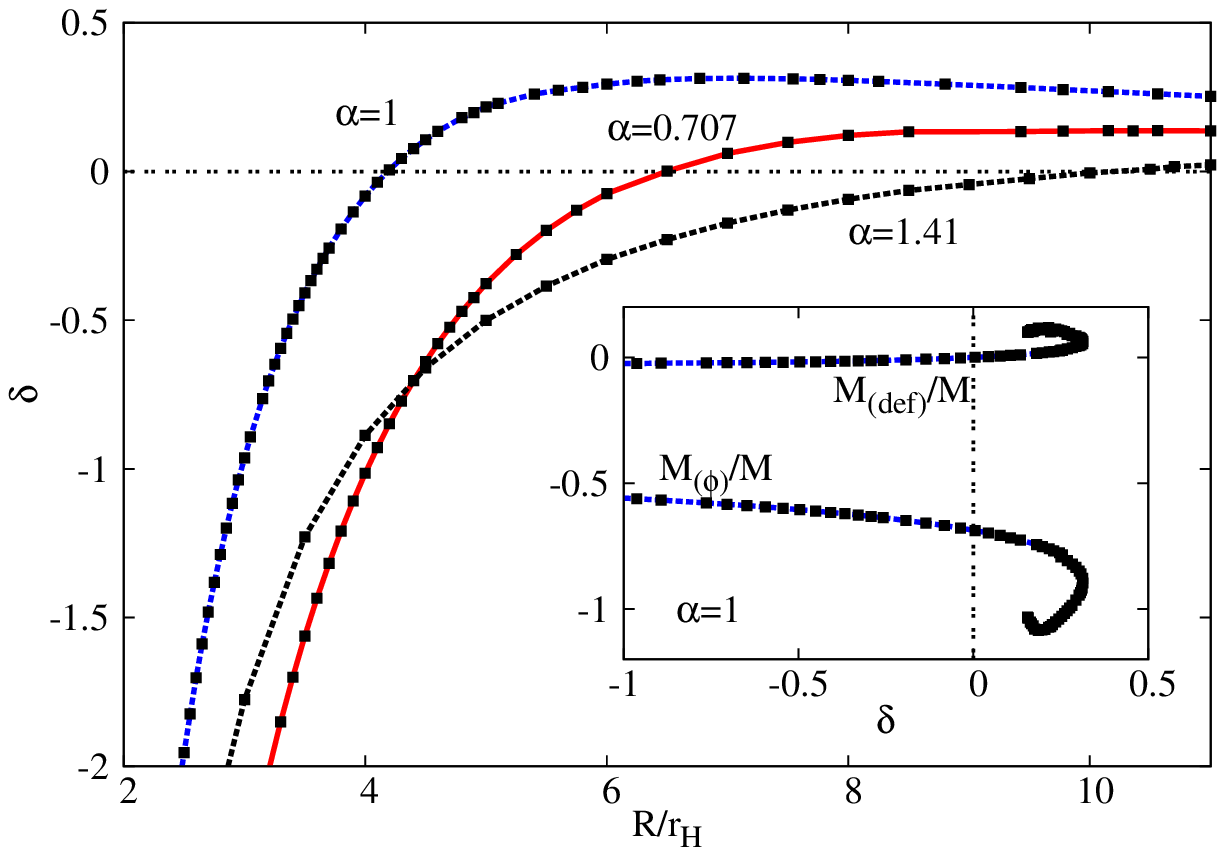}}
\hss}
\caption{\small 
The conical defect
$\delta$
 is shown as a function of the coupling parameter $\alpha$ 
and as a function of the ratio $R/r_H$ (with $R$ the radius of the ring and $r_H$ the event horizon radius).
In both cases, one notices the existence of balanced configurations ($\delta=0$).
The inset shows the ratio between the total mass associated with the conical defect $M_{({\rm def})}$
and the ADM mass $M$ as a function of $\delta$
(and the same for the mass $M_{(\phi)}$ stored in the scalar field). 
}
\label{data-ring}
\end{figure}

\subsection{The black rings}
 
Starting again with the probe limit, 
we have solved the equation for $\phi$ 
in a vacuum BR background as given by (\ref{BR-vacuum}).
For a given horizon radius $r_H$,
the solutions were found up to a maximal value of the radius $R$,
where the errors become large. 
The profile of a typical solution is shown in Figure \ref{probe-BR}.
One can see both the scalar field and the energy density
possess a non-trivial angular dependence,
with a maximum located at the horizon for some intermediate value of $\theta$.

 {\small \hspace*{3.cm}{\it  } }
\begin{figure}[t!]
\hbox to\linewidth{\hss%
	\resizebox{9cm}{7cm}{\includegraphics{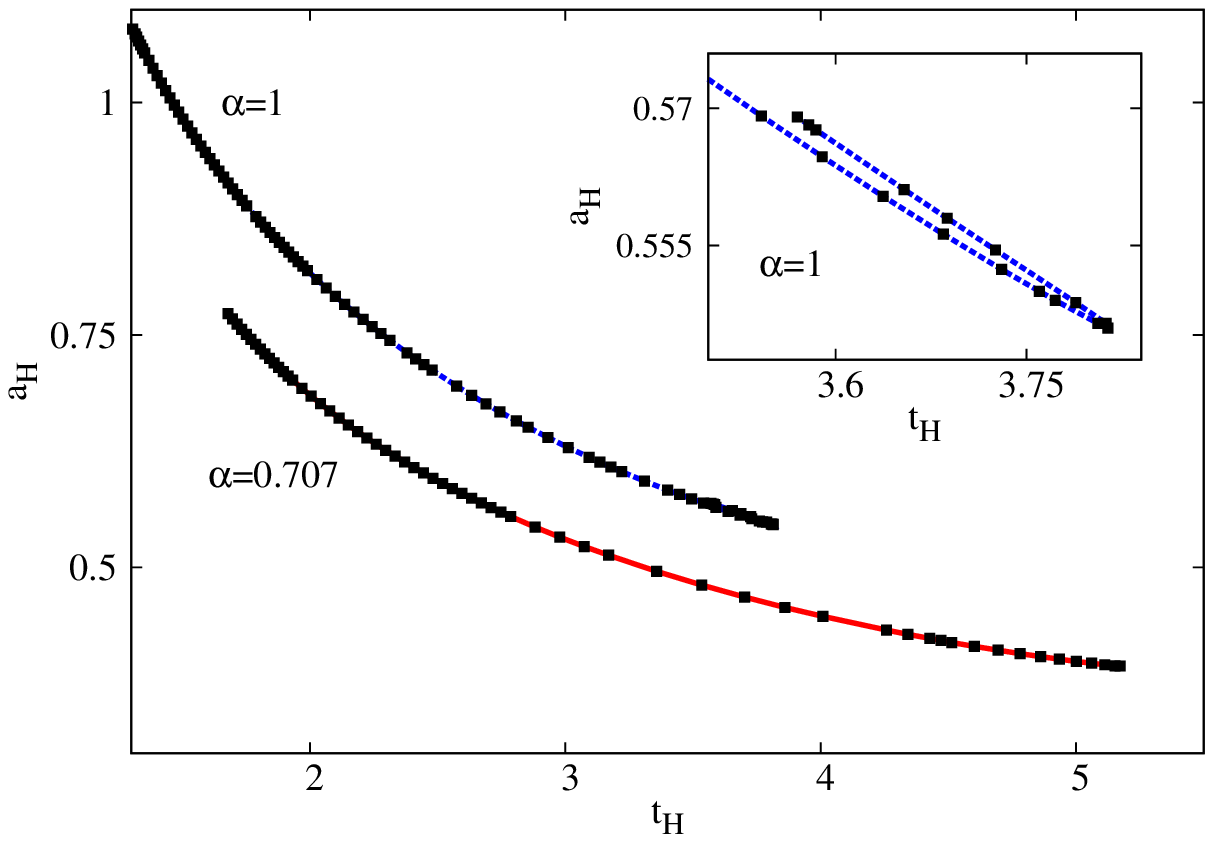}}
	\resizebox{9cm}{7cm}{\includegraphics{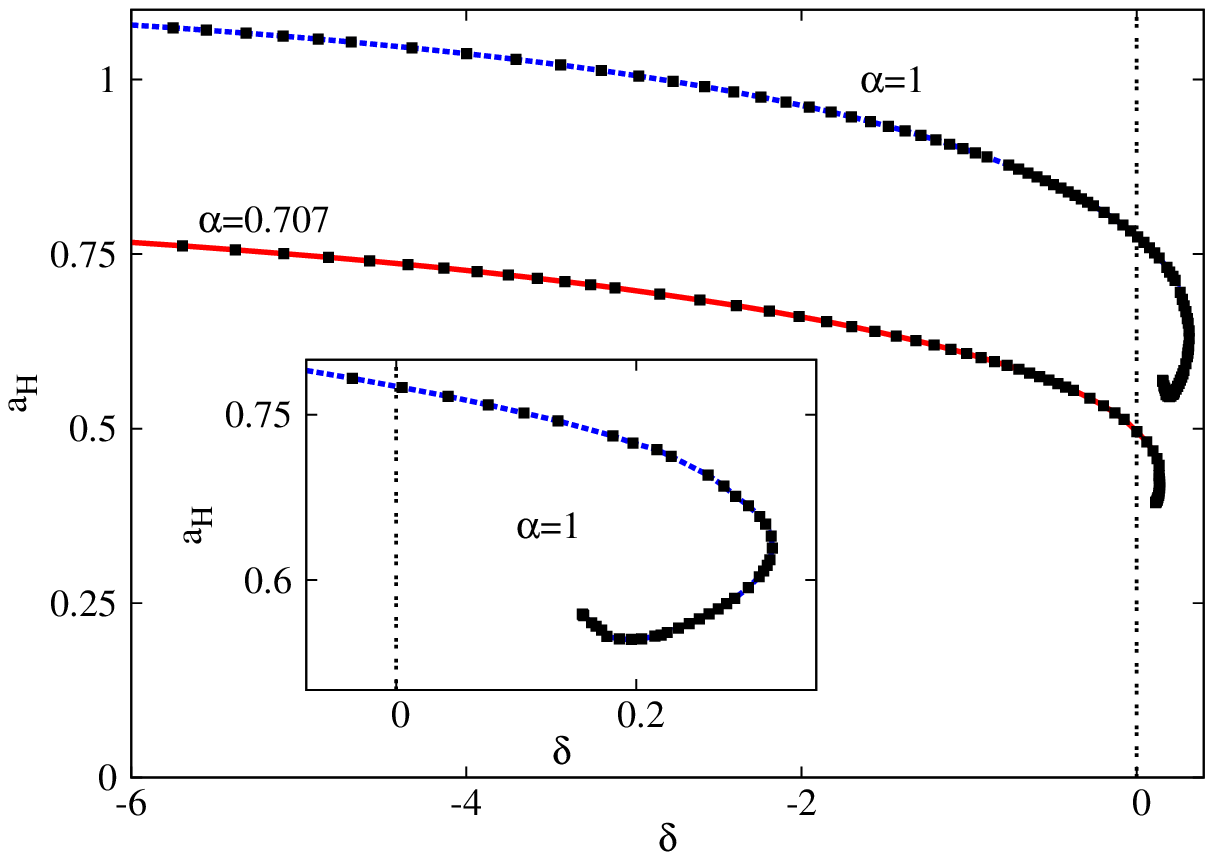}}
\hss}
\caption{\small 
The reduced area 
$a_H$
is shown as a function of reduced temperature 
$t_H$
and conical defect
$\delta$
 for given values of $\alpha$.
}
\label{data-ring2}
\end{figure}

The backreacting generalizations  of these solutions are found again by 
increasing from zero the parameter $\alpha$. 
As in the spherical case, 
this results in a  complicated branch structure, and 
more than one solution may exist for the same input parameters
$(\alpha; r_H,R)$.
The BRs are regular on and outside the horizon and
show no sign of a singular behaviour.
However, 
as expected, 
the generic configurations possess a conical singularity. 
As one can see from the boundary conditions (\ref{bc-inf}),
in this work we have 
chosen\footnote{It is also possible to work 
with the conical singularity stretching
towards the boundary, in which case the spacetime
will not be asymptotically flat.} 
to locate the conical singularity 
at  $\theta=0$, $r_H<r<R$, where we find a conical singularity, as measured by the parameter
\begin{eqnarray}
\label{delta}
\delta= 2\pi (1-\lim_{\theta\to 0}\frac{f_2}{\theta^2 r^2 f_1 }) \neq 0~.
\end{eqnarray}
(Note that a vacuum BR has
$
\delta= - {4\pi r_H^2}/{(R^2-r_H^2)}<0
$, with $\delta$ diverging in the Schwarzschild-Tangerlini limit.)
This can be interpreted as a disk
preventing the collapse of the configurations.
Although the presence of a conical singularity is an undesirable feature,
it has been argued in 
\cite{Herdeiro:2009vd}, 
\cite{Herdeiro:2010aq},
that such asymptotically flat black objects still
admit a thermodynamical description (see also \cite{Astefanesei:2009mc}). 
Moreover, when working with the appropriate set
of thermodynamical variables, the Bekenstein-Hawking law still holds, 
while the parameter $\delta$
enters the first law of thermodynamics, corresponding to a pressure term $P$,
with the conjugate extensive variable  ${\cal A}$,  
\begin{eqnarray}
\label{P}
P=-\frac{\delta}{8\pi}~~{\rm and }~~{\cal A}= Area~T_H,
\end{eqnarray}
 where $Area$ is the space-time area of the conical singularity's world-volume,
as computed from the line-element
 $
 d\sigma^2=-f_0dt^2+f_1dr^2+f_3d\varphi^2.
 $
 For the line-element (\ref{metric}),  
one finds
\begin{eqnarray}
\label{Area}
{\cal A}= 2\pi \int_{r_H}^R dr \sqrt{f_0f_1 f_3 }\bigg|_{\theta=0}.
\end{eqnarray} 
Then the total mass-energy  associated with the conical defect is  \cite{Kleihaus:2010pr}:
\begin{eqnarray}
\label{Mdef}
M_{({\rm def})}=-P {\cal A}.
 \end{eqnarray}  

As expected, the (absolute) value of 
the conical excess $\delta$
decreases as $\alpha$
is increased ($i.e.$ allowing for a larger $M_{(\phi)}$ contribution to the total mass).
Therefore, for a BR set with fixed
horizon and ring radii $(r_H,R)$, a balanced configuration
is achieved for a critical value of $\alpha$.
Further increasing $\alpha$
results in a configurations with a 
conical excess $\delta>0$, see Figure \ref{data-ring} (left panel).

When considering instead a model with a fixed coupling constant  $\alpha>0$ 
and varying  the size of the ring,
this also results in the 
existence of a critical balanced configuration.
The results for several value of $\alpha$
are shown in Figure \ref{data-ring} (right panel).
One can see that the (absolute value of the) total mass associated with the defect
$M_{({\rm def})}$
is always small as compared to the ADM mass $M$, 
while  the mass associated with the scalar field $M_{(\phi)}$ takes negative values,
and dominates over the horizon mass for $\delta>0$.
 
The limit $R\to r_H$
of the solutions appears to be similar to the vacuum case,
a BH solution with spherical horizon topology 
being approached (although this limit is difficult to study in our numerical scheme).
Rather surprising, no arbitrarily large BRs were found for the cases investigated so far. 
Instead, as seen in  Figures \ref{data-ring}, \ref{data-ring2},
the solutions stop to exist for a maximal value of $\delta$,  
with a backbending and the occurrence of a secondary branch. 
  However, clarifying the critical behaviour,
 together with a systematic investigation 
of the parameter space of solutions is beyond the purposes of this work.  

\section{Further remarks}
%
%
The known five dimensional, static 
black rings (BRs) in a Minkowski spacetime background are plagued
by conical singularities.
As shown in this work, this pathology can be cured at the price of 
coupling Einstein gravity with a {\it 'phantom'} scalar field.
In such a model, when
fixing the coupling constants, 
balanced solutions were
shown to exist for critical radii of a BR.

The spinning, balanced, Emparan-Reall BRs are known to possess 
higher dimensional generalizations
\cite{Kleihaus:2012xh},
\cite{Dias:2014cia} 
(although a closed form solution is still missing).
Moreover, when increasing  the number $d$ of spacetime dimensions,
a plethora of other black objects with various event horizon topologies are found
(for a review, see  
\cite{Emparan:2008eg}).
%
While the unbalanced $d>5$ BRs appear to be singular,
(at least) the solutions with 
a $S^2\times S^{d-4}$ horizon topology
possess a well defined  static limit,
with conical singularities only 
\cite{Kleihaus:2009wh},
\cite{Kleihaus:2014pha}.
The results in this work 
suggest that these {\it ringoids} 
achieve balance when including
a phantom field in the model.

\medskip

Moreover, one can speculate that the same mechanism could allow for the
existence of four dimensional BRs.
The results of various theorems
excluding a non-spherical topology of the horizon
\cite{HE} 
would be circumvented 
for  an exotic matter content 
violating the 
energy conditions (see 
\cite{Iizuka:2007sk}
for some speculations in this direction).

In fact, following the 
approach in the Introduction, one can easily write
a line element describing a four dimensional, asymptotically flat BH  
which is regular on and outside an horizon 
of $S^1\times S^1$ topology.
Although this geometry does not solve any obvious field theory model,  
it may give an idea about the properties of a four dimensional BR  solution.
%
%
For concreteness, let us consider the following metric:
\begin{eqnarray}
\label{metric1}
ds^2=\frac{R^2}{(x-y)^2}
\left[
\frac{dx^2}{1-x^2} 
+\frac{(1+\lambda x)^2}{H(x,y)} 
                                      \left ( 
																		\frac{1}{1+\lambda y}	\frac{dy^2}{y^2-1}
																			+ \frac{y^2-1}{1-\lambda}d\varphi^2
																		 \right )
\right]
-\frac{1+\lambda y}{H(x,y)}dt^2~,
\end{eqnarray}
where $R$, $\lambda$ are free parameters
(with $R>0$
and $0<\lambda<1$),
while
$x,y$ are toroidal coordinates, with 
 $-\infty\leq y\leq -1$, $-1\leq x\leq 1$,  the  asymptotic infinity being at $x\to y \to -1$.
Also, $H(x,y)$ is a smooth, strictly positive function
(with smooth derivatives as well),
 which controls the far field behaviour of the geometry.
Then one can easily verify the absence of a
conical singularity for the line-element (\ref{metric1}),
the periodicity of $\varphi$ being $2\pi$, as usual.

The line element (\ref{metric1})
possesses an event horizon located at 
$
y=-1/{\lambda}<-1,
$
the  metric of its spatial cross-section being
\begin{eqnarray}
\label{metric1h}
d\sigma^2=R^2
\left(
\frac{\lambda^2}{(1+\lambda x)^2(1-x^2)} dx^2
+  \frac{1+\lambda }{H(x,-1/\lambda)}d\varphi^2 
\right).
\end{eqnarray}
This horizon has an $S^1\times S^1$ topology, 
as results $e.g.$
from the fact that
 its Euler characteristic vanishes.
%
Also,
the  Hawking temperature and the event horizon area corresponding to the metric (\ref{metric1})  are
well defined, with
\begin{eqnarray}
T_H=\frac{\sqrt{1-\lambda^2}}{4\pi R \lambda},~~
A_H=2\pi  R^2 \lambda \sqrt{1+\lambda} 
\int_{-1}^1 dx 
\left(
(1+\lambda x)\sqrt{(1-x^2) H(x,-1/\lambda)}
\right)^{-1}.
\end{eqnarray} 

The line-element (\ref{metric1})  
has an associated energy-momentum tensor whose 
 nonzero components (as found from 
the Einstein equations) are
$T_{xx}$,
$T_{xy}$,
$T_{yy},$
$T_{\varphi \varphi}$
and 
$T_{tt}$,
whose explicit form depend on the choice of $H(x,y)$. 
The simplest expression of this function 
compatible with regularity and 
the required asymptotic behaviour is
\begin{eqnarray}
\label{G}
H(x,y)=(1-\lambda) 
\left(
1+\nu \sqrt{x-y}
\right),
\end{eqnarray}
with $\nu>0$ a free parameter.
%
Then the resulting line-element
appears to be regular and free of pathologies on and outside the horizon.
For example. the power series expansion of various quantities
(like Kretschmann scalar, $R$ and $E_\mu^\nu$)
at $y=-1/\lambda$, $y=-1$
and $x=\pm 1$ is free of singularities.
Also,
smooth profiles are found
when plotting the same quantities for various choices
of the parameters $\lambda,~\nu$
 (with $R=1$ without any loss of generality). 

In the study of the far field expression of various quantities,
we consider the following 
coordinate transformation
\begin{eqnarray}
x=-\frac{r^2-R^2}{\sqrt{  (r^2-R^2)^2+4 r^2 R^2 \cos^2 \theta }},~~
y= -\frac{r^2+R^2}{\sqrt{ (r^2-R^2)^2+4 r^2 R^2 \cos^2 \theta }},
\end{eqnarray}
with $r$, $\theta$ 
possessing (for large $r$)
the usual interpretation, 
and 
$0\leq \theta\leq \pi$.
Then 
the Minkowski spacetime is recovered as $r\to \infty$, and one finds $e.g.$
\begin{eqnarray}
g_{tt}=-1+\frac{\sqrt{2} \nu R}{r}+O(1/r^2)+\dots,
\end{eqnarray}
which implies an ADM mass  
$M=\frac{\nu R}{\sqrt{2} G}>0.$ 
However, one can easily show that, as expected,
the energy density of the matter source, $\rho=-T_t^t=-E_t^t/(8\pi G)$,
takes negative value for some region  on and outside the horizon.

The basic results above hold as well for other choices of 
the function  
$H(x,y)$, and also for several generalizations of the line-element (\ref{metric1}) we have considered.
In all cases, we were not able to identify a field theory source for the energy-momentum tensor
compatible with such metrics.
However,  (\ref{metric1}) (or another version of it) could be useful as providing a $background$ 
geometry in a numerical attempt to construct four dimensional BRs
for a model 
with a matter source allowing for negative energy densities, in particular with a phantom scalar field.

\section*{Acknowledgements}
The authors thank D. Astefanesei, P. Cunha and P. Nedkova
for useful remarks on a draft of this paper.
B.K. and J.K. gratefully
acknowledge support by the DFG Research Training Group 1620 Models of Gravity.
The  work of E.R. is supported by the Funda\c{c}\~ao para a Ci\^encia e a Tecnologia (FCT) project UID/MAT/04106/2019 (CIDMA), 
by CENTRA (FCT) strategic project UID/FIS/00099/2013, 
by national funds (OE), through FCT, I.P., in the scope of the framework contract foreseen in the numbers 4, 5 and 6
of the article 23, of the Decree-Law 57/2016, of August 29,
changed by Law 57/2017, of July 19, and also by the project PTDC/FIS-OUT/28407/2017.  
 This work has further been supported by  the  European  Union's  Horizon  2020  research  
and  innovation  (RISE) programmes H2020-MSCA-RISE-2015
Grant No.~StronGrHEP-690904 and H2020-MSCA-RISE-2017 Grant No.~FunFiCO-777740.
E.R.
gratefully acknowledges the support of the Alexander von Humboldt Foundation.
The authors would like to acknowledge networking support by the
COST Action CA16104. 
Computations were performed
at the BLAFIS cluster, in Aveiro University.

 \begin{small}


 \end{small}

 \end{document}